\documentclass[letterpaper, 10 pt, conference]{ieeeconf}  

\IEEEoverridecommandlockouts                              

\usepackage{color}
\usepackage{graphics} 
\usepackage{epsfig} 
\usepackage{subfigure}
\usepackage{times} 
\usepackage{amsmath} 
\usepackage{amssymb}  

\newcommand{\R}{\mathbb{R}}
\newtheorem{theorem}{Theorem}[section]

\newtheorem{definition}{Definition}[section]
\newtheorem{lemma}[definition]{Lemma}

\newtheorem{remarkth}[definition]{Remark}

\usepackage{amssymb}

\newcommand{\proa}{A^*G \mbox{$\;$}_{\tau^*} \kern-3pt\times_\alpha
G \mbox{$\;$}_\beta \kern-3pt\times_{\tau^*} A^*G}

\title{\LARGE \bf
Variational collision and obstacle avoidance of multi-agent systems on Riemannian manifolds}
\author{Rama Seshan Chandrasekaran, Leonardo J. Colombo, Margarida Camarinha, Ravi Banavar, Anthony Bloch 
\thanks{Rama Seshan (ee17d402@smail.iitm.ac.in) is with Indian Institute of Technology Madras, Chennai, India.  L.
Colombo (leo.colombo@icmat.es) is with Instituto de Ciencias Matem\'aticas
(ICMAT), Calle Nicolas Cabrera 13-15, 28049, Madrid, Spain. M. Camarinha ( mmlsc@mat.uc.pt) is with CMUC, Department of Mathematics, University of Coimbra, 3001-501 Coimbra, Portugal. R. N. Banavar (banavar@iitb.ac.in) is with the Systems and Control Enginerring, Indian Institute of Technology Bombay, Mumbai, Maharashtra
400076, India. A. Bloch (abloch@umich.edu) is with Department of Mathematics, University of Michigan, 530 Church St. Ann Arbor, 48109, Michigan, USA.
The research of A. Bloch was supported by NSF grant DMS-1613819 and AFOSR grant 17RT0038. The research of M. Camarinha was partially supported by the Centre for Mathematics of  the University of Coimbra -- UID/MAT/00324/2019, funded by the Portuguese Government through FCT/MEC and co-funded by the European Regional Development Fund through the Partnership Agreement PT2020. L. Colombo was supported by This work was partially supported by I-Link Project (Ref: linkA20079) from CSIC; Ministerio de Economia, Industria y Competitividad (MINEICO, Spain) under grant MTM2016-76702-P; ``Severo Ochoa Programme for Centres of Excellence'' in R$\&$D (SEV-2015-0554).The project that gave rise to these
results received the support of a fellowship from ``la Caixa' Foundation (ID 100010434). The fellowship
code is LCF/BQ/PI19/11690016. }} 

\begin{document}

\maketitle
\thispagestyle{empty}
\pagestyle{empty}

\begin{abstract}
In this paper we study a path planning problem from a variational approach to collision and obstacle avoidance for multi-agent systems evolving on a Riemannian manifold. The problem consists of finding non-intersecting trajectories between the agent and prescribed obstacles on the workspace, among a set of admissible curves, to reach a specified configuration, based on minimizing an energy functional that depends on the velocity, covariant acceleration and an artificial potential function used to prevent collision with the obstacles and among the agents. We apply the results to examples of a planar rigid body, and collision and obstacle avoidance for agents evolving on a sphere.  
 \end{abstract}

\section{Introduction}
Cooperation and coordination of multi-agent systems has received a great deal of attention due to its widely practical applications in both civilian and military areas \cite{Jorgebook}, \cite{MMbook}. These applications include cooperative control of unmanned aerial vehicles and satellite clusters, flocking, formation control, and control of sensor networks for exploration, surveillance, rescue and missions, video capture in sport events, transportation, etc. Among these tasks of cooperation and coordination control fields,  the search for optimal motion plans can be computationally very expensive, particularly as the number of robots or degrees of freedom of the system gets large. For this reason, we pursue in the work an easily computable motion planning strategy, based on geometric techniques and variational calculus, to avoid collision between agents and static obstacles in the workspace. 

Calculus of variations on Riemannian manifolds \cite{Milnor} has been exploited in the past for various applications. In Crouch and Silva Leite \cite{CroSil:91}, \cite{CroSil:95} the authors have used it
to develop a theory of generalized cubic polynomials for
dynamic interpolation problems on Riemannian manifolds, in particular on compact connected Lie groups endowed with a bi-invariant metric.
More recently, Bloch, Camarinha and Colombo \cite{BlCaCoCDC}, \cite{BlCaCoCDC2}, \cite{BCC}  have
used these variational methods to solve obstacle avoidance
problems on Riemannian manifolds and interpolation problems.

 In this paper, as a continuation of the work \cite{Assif}, we address the problem of finding necessary conditions for optimal trajectories of multiple agents on a Riemannian manifold that seek to achieve a specified configuration while avoiding collisions among themselves and static obstacles on the configuration space.

Specifically, the problem studied in this paper consists of finding non-intersecting trajectories of a given number of agents, among a set of admissible curves, to
reach a specified configuration and minimizing an energy functional that depends on the velocity, covariant acceleration and an artificial potential function used to avoid obstacles and collisions between the agents. To solve the problem, we employ techniques from calculus
of variations on Riemannian manifolds taking into account
that the problem under study can be seen as a higher-order
variational problem \cite{blochcrouch2}, \cite{MargaridaThesis}, \cite{CMdD}, \cite{BlochHussein}.


The structure of the paper is as follow. We start in Section II by
introducing the geometric structures on a Riemannian manifold
that we will use together with admissible variation of curves
and vector fields for the variational problem. Next, in Section III, we introduce variational collision and obstacle avoidance problems on Riemannian
manifolds and first-order derive necessary conditions for optimality. We apply the result to
obstacle and collision avoidance problems for a planar rigid body and agents evolving on a sphere in Section IV. Final comments and ongoing work are discussed at the end of the paper.
\section{Preliminaries on Riemannian Geometry}

Let $M$ be a smooth $ (C^\infty) $ Riemannian manifold with Riemannian metric denoted by $ <\cdot,\cdot>\ : T_{x}M \ \times \ T_{x}M \rightarrow \mathbb{R}$ at each point $x \in M $, where $T_{x}M$ is the tangent space of $M$ at $x$.  The length of a tangent vector is determined by its norm,
$||v_x||=\langle v_x,v_x\rangle^{1/2}$ with $v_x\in T_xM$, for each point $x\in M$.


A \textit{Riemannian affine connection} $\nabla$ on $M$, is a map that assigns to any two smooth vector fields $X$ and $Y$ on $M$, a new vector field, $\nabla_{X}Y$, called the \textit{covariant derivative of $Y$ with respect to $X$}, satisfying $\nabla_{fX}Y=f\nabla_XY,\hbox{ and } \nabla_X(fY)=X(f)Y+f\nabla_XY$, for all vector fields $X,Y\in\mathfrak{X}(M)$ and $f\in\mathcal{C}^{\infty}(M)$, where $\mathfrak{X}(M)$ denotes the set of vector fields on $M$. For the properties of $\nabla$, see for instance \cite{Boothby, bookBullo,Milnor}.

Consider a vector field $W$ along a curve $x(t)$ on $M$. For $n\leq 1$, the $n$-th order covariant derivative of $W$ along $x(t)$ is denoted by $\displaystyle{\frac{D^n W}{dt^n}}$ and we will denote by $\displaystyle{\frac{D^{n+1} x}{dt^{n+1}}}$, the $n$-th order covariant derivative along $x$ of the velocity vector field of $x$.

Given vector fields $X$, $Y$
and $Z$ on $M$, the vector field $R(X,Y)Z$ given by \begin{equation}\label{eq:CurvatureTensorDefinition}
R(X,Y)Z=\nabla_{X}\nabla_{Y}Z-\nabla_{Y}\nabla_{X}Z-\nabla_{[X,Y]}Z
\end{equation}  defines the \textit{curvature tensor} on $M$, where $[X,Y]$ denotes the \textit{Lie bracket} of the vector fields $X$ and $Y$. $R$ is trilinear in $X$, $Y$ and $Z$ and a tensor of type $(1,3)$. For vector fields $X,Y,Z,W$ on $M$ the curvature tensor satisfies (see \cite{Milnor}, p. 53)
\begin{equation}\label{curvformula}\langle R(X,Y)Z,W\rangle=\langle R(W,Z)Y,X\rangle.\end{equation}

Let $S$ be a submanifold of $M$ and $\Omega$ be the set of all $C^{1}$ piecewise smooth curves $x : [0,T] \ \rightarrow M$ such that $x(0)$, $\frac{dx}{dt}(0)$ and $x(T)$ are fixed, with $x(T)\in S$ and $\frac{dx}{dt}(T) \in T_{x(T)}S$. The set $\Omega$ is called the \textit{admissible set}. For the class of curves in $\Omega$ we introduce the $C^1$ piecewise smooth \textit{one parameter admissible variation} of a curve $x \in \Omega$ by $\alpha : (-\epsilon,\epsilon) \times [0,T] \rightarrow M ;(r,t) \mapsto\alpha(r,t) = \alpha_r(t)$ that satisfy $\alpha_{0} = x$ and $\alpha_r \in \Omega$, for each $r \in (-\epsilon, \epsilon) $.

 The \textit{variational vector field} associated to an admissible variation $\alpha$ is a $C^1$-piecewise smooth vector  field $X$ along $x$ defined by $\displaystyle{X(t) = \frac{D}{\partial r}\Big{|}_{r=0}\alpha(r,t) \in T_{x(t)}\Omega}$ verifying the boundary conditions
\begin{equation*}
X(0)=0,\, X(T) = 0,\,
\frac{DX}{dt}(0)=0, \, \frac{DX}{dt}(T) \in T_{x(T)}S,
\end{equation*} where the tangent space of $\Omega$ at $x$ is the vector space $T_{x}\Omega$ of all $C^1$ piecewise smooth vector fields $X$ along $x$ verifying the former boundary conditions. 

\begin{lemma}[\cite{Milnor}, p.$52$]\label{curvature_lemma}
The one parameter variation satisfies
\begin{equation*}
\frac{D}{\partial r}\frac{D^2\alpha}{\partial t^2} = \frac{D^2}{\partial t^2}\frac{\partial \alpha}{\partial r} + R \Big( \frac{\partial \alpha}{\partial r},\frac{\partial \alpha}{\partial t} \Big)\frac{\partial \alpha}{\partial t}.
\end{equation*}
\end{lemma}

\vspace{.2cm}

\section{The variational collision and obstacle avoidance problem on Riemannian manifolds}

\label{S:3}
 Let $n$ and $k$ be natural numbers and $T$ be a positive real number. Consider $n$ agents evolving on $M$, a Riemannian manifold with $\dim(M)=m$. Denote by $(p_{0}^{i}, v_{0}^{i})$, for $i=1,2...,n$,  points in $TM$ corresponding to the initial positions and velocities of the agents.

Consider the set $\Omega_{i}$ of all $C^1$-piecewise smooth curves $x_{i} : [0,T] \rightarrow M$ verifying the boundary conditions
\begin{equation*}
x_i(0) = p_{0}^{i}, \, \frac{dx_i}{dt}(0) = v_{0}^{i},\, x_i(T) = p_T^i \in S, \, \frac{dx_i}{dt}(T) \in T_{x_i(T)}S
\end{equation*} that is, each agent is initially at a fixed position and velocity, and required to reach in a fixed time, a specified position on the submanifold $S$ with its velocity tangent to $S$ at the specified point.

The problem studied in this work consists on designing a trajectory for each agent, satisfying the above boundary conditions and avoiding a set of prescribed static obstacles in the workspace together with ensuring collision avoidance between the agents. The path planning is designed by considering a cost functional which is defined on the set of $C^1$ piecewise smooth trajectories, verifying the above boundary conditions, such that the trajectory that minimizes the defined cost functional will be a feasible solution avoiding collision between the agents and each agent will avoid the obstacles.

Figure $1$ gives a sketch of the situation and the nomenclature used in this paper is given in Table \ref{tab:table1}.

\begin{figure}[h]\label{fig1}
 \includegraphics[scale=0.45]{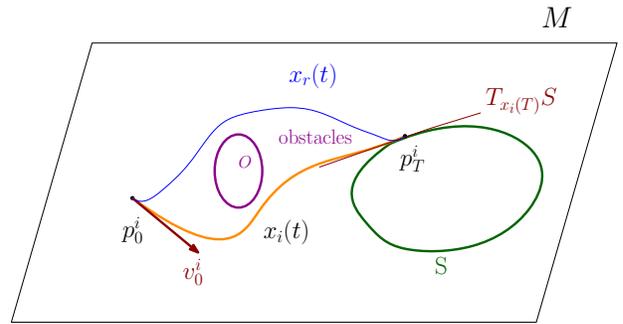}
 \caption{Sketch of the situation studied in this work. $x_i(t)$ and $x_r(t)$ denotes trajectories for the $i^{th}$ and $r^{th}$ agents, respectively.} 
\end{figure}

\begin{table}[h!]
  \begin{center}
    \caption{Nomenclature}
    \label{tab:table1}
    \begin{tabular}{l|c} 
    \hline 
      \textbf{Symbol} & \textbf{Description} \\
      \hline
     $M$ & Riemannian manifold. \\
    $S$ & Riemannian submanifold. \\
	$n$ & Number of agents\\
	$T$ & Final time\\
	$J$ & Cost functional\\
	$V$ & Artificial Potential function \\
	$x_i$ & Trajectory of the $i^{th}$ agent\\
	$m$ & Dimension of $M$ \\
	$s$ & Dimension of $S$ \\
	$\frac{D}{dt}$  & Covariant derivative \\
	$\alpha$ & One parameter smooth variation of $x_i$ \\
	$t_{0_i},t_{1_i},...,t_{l_i}$ & Partition of $[0,T]$ for the $i^{th}$ agent \\
	$d_M(x,y)$ & Geodesic distance between $x,y\in M$ \\
	$\alpha$  & Quantity of static obstacles\\
	$B_{\alpha}$ & Smooth real valued function  \\ & describing the $\alpha$ static obstacles 
    \end{tabular}
  \end{center}
\end{table}


Define the cost functional $J$ on $\tilde{\Omega} = \Omega_{1}\times$...$\times\Omega_{n}$ as
\begin{align*}
J(x_1, x_2, ..., x_n) =&\frac{1}{2} \int_{0}^{T}\sum_{i=1}^{n} \Big( \Big{\|} \frac{D^2x_i}{dt^2}(t) \Big{\|}^2 + \kappa\Big{\|}\frac{dx_i}{dt}(t)\Big{\|}^2\\&+  V(x_1, x_2, ..., x_n)\Big) dt.
\end{align*}

\vspace{.2cm}

The cost functional $J$  is constructed as a combination of the square magnitudes for covariant acceleration, square magnitudes of velocity regulated by a parameter $\kappa$, both for the individual trajectories of the agents, and an artificial potential function $V$ that penalizes collisions between the agents and also with prescribed static obstacles. The obstacles are described as the zero level surface of a know scalar valued analytic function (see, e.g., \cite{Khatib1986}, \cite{K88}, \cite{Koditschek1990}).


The potential function $V$ is an artificial potential field-based function
represented by a force inducing an artificial repulsion from the surface of the obstacle as. We use the approach introduced by Khatib \cite{Khatib1986}, which consists of using a local inverse potential field going to infinity as the inverse square of a known scalar valued analytic function near the obstacle and collision between the agents, and decaying to zero at some positive level surface far away from the obstacle and when agents arte not close to each other.


Let $d_M$ denote the distance function induced by the Riemannain metric on $M$, and $\tau$, $\sigma$ be positive constants. Let the obstacle be the zero sublevel set of a smooth function $B$ in $M$. i.e. $O=\{x \in M | B(x)\leq 0\}$. Then the artificial potential function takes the (non-unique) form 

$$V(x_1,..,x_n)=\sum_{i=1}^n\left(\sum_{k=1}^{\alpha}\frac{\tau}{B_{\alpha}(x_i)} + \frac{1}{2}\sum_{j=1}^n\frac{\sigma}{d_M(x_i,x_j)}\right)$$ where $\alpha$ denote the quantity of obstacle. Note that the factor $\frac{1}{2}$ in the expression for $V$ comes from the fact that $d_M(x_i,x_j)=d_M(x_j,x_i)$. 

\textbf{Problem:} The variational collision and obstacle avoidance problem consists of minimizing the functional $J$ among $\tilde{\Omega}$. 

\vspace{.2cm}

In order to minimize the functional $J$ among the set $\tilde{\Omega}$ we want to find curves $x \in \tilde{\Omega}$ such that $J(x) \leq J(\tilde{x})$, for all admissible curves $\tilde{x}$ in a $C^1$ neighborhood of $x$.
\vspace{.1cm}

\begin{theorem}\label{theorem1}
Let $x_i \in \Omega_i$. If $\alpha$ is an admissible variation of $x_i$ with variational vector field $X_i$, then
\begin{align*}
&\frac{d}{dr}J(\alpha(r))\Big{|}_{r=0}= \int_{0}^{T}\sum_{i=1}^n \Bigg( \Big  \langle X_i,\frac{D^4 x_i}{dt^4}- \kappa\frac{D^2x_i}{dt^2} \\&+ R \Big(\frac{D^2 x_i}{dt^2}, \frac{dx_i}{dt} \Big)\frac{dx_i}{dt}+\mbox{grad}_i V(x_1, x_2, ..., x_n)  \Big \rangle \Bigg) \ dt \\ +&\sum_{i=1}^n \sum_{j_i=1}^{l_i} \Big{[}\Big\langle\frac{DX_{i}}{dt},\frac{D^2x_{i}}{dt^2} \Big \rangle+\Big\langle X_i, \kappa \frac{dx_i}{dt} - \frac{D^3x_i}{dt^3} \Big \rangle\Big{]}^{t_{j_i}^{-}}_{t_{j_{i-1}}^{+}},
\end{align*} where $0=t_{0_i}<....<t_{l_i}=T$ is a partition of $[0,T]$ such that $x_i$ is smooth on each $[t_{j_i},t_{j_i+1}]$, $i=1,\ldots,n$ and $j_i=0,\ldots,l_{i-1}$, and where $\mbox{grad}_i V$ denotes the gradient of $V$ with respect to its $i^{th}$ argument.
\end{theorem}

\vspace{.2cm}

\textit{Proof:} If $\alpha$ is an admissible variation of $x_i \in \Omega_i$ with variational vector field $X_i$, then
\begin{align*}
 \frac{d}{dr}J(\alpha_r) =& \int_0^T \sum_{i=1}^n \Bigg( \Big \langle \frac{D}{\partial r} \frac{D^2 \alpha}{\partial t^2}, \frac{D^2 \alpha}{\partial t^2}\Big \rangle + \kappa\Big \langle\frac{D^2 \alpha}{\partial r \partial t}, \frac{\partial \alpha}{\partial t} \Big \rangle \\&+ \frac{\partial}{\partial r}  V(x_1, x_2, ...,x_{i-1}, \alpha, x_{i-1}, ..., x_n) \Bigg)dt.
\end{align*}

By lemma \ref{curvature_lemma}, the property of curvature tensor (2), and the previous equation, we get

\begin{align*}
\frac{d}{dr}J(\alpha_r) = &\int_0^T\sum_{i=1}^n \Bigg( \Big \langle \frac{D^2}{dt^2}\frac{\partial \alpha}{\partial r}, \frac{D^2 \alpha}{dt^2}\Big \rangle  \\&+ \Big \langle R\Big(\frac{\partial \alpha}{\partial r}, \frac{\partial \alpha}{\partial t}\Big) \frac{\partial \alpha}{\partial t}, \frac{D^2 \alpha}{\partial t^2} \Big \rangle +  \kappa\Big \langle\frac{D^2 \alpha}{\partial r \partial t}, \frac{\partial \alpha}{\partial t} \Big \rangle \\ +& \Big \langle  \frac{\partial\alpha}{\partial r}, \mbox{grad}_i V(x_1, x_2, ..., x_n)  \Big \rangle \Bigg)dt.
\end{align*}

Integrating the first term by parts twice, the third term once, and applying Lemma \ref{curvature_lemma} to the second term, we obtain that
\begin{align*}
\frac{d}{dr}J(\alpha_r) =& \int_0^T \sum_{i=1}^n\Bigg( \Big \langle \frac{\partial \alpha}{\partial r}, \frac{D^4\alpha}{dt^4} + R \Big( \frac{D^2\alpha}{dt^2}, \frac{\partial \alpha}{\partial t} \Big) \frac{\partial \alpha}{\partial t} \\&- \kappa\frac{D^2\alpha}{dt^2} +\mbox{grad}_i V(x_1, x_2, ..., x_n)\Big \rangle \Bigg) dt \\ +&\sum_{i=1}^n \sum_{j_i=1}^{l_i} \Big{[} \Big \langle \frac{D}{dt}\frac{\partial \alpha}{\partial r},\frac{D^2\alpha}{dt^2} \Big \rangle \\&+ \Big \langle \frac{\partial \alpha}{\partial r}, \kappa \frac{\partial \alpha}{\partial t} - \frac{D^3\alpha}{\partial t^3} \Big \rangle\Big{]}^{t_{j_i}^{-}}_{t_{j_i-1}^{+}}
\end{align*}
where the interval $[0,T]$ is partitioned as $ 0 = t_{0_i} < t_{2_i} < ...< t_{l_i} = T $ such that in each subinterval $x_i$ is smooth. 

Taking $r = 0$ in the last equation, 
\begin{align*}
 \frac{d}{dr}J(\alpha_r)\Big{|}_{r=0} =& \int_0^T \sum_{i=1}^n\Bigg( \Big \langle X_i, \frac{D^4x_i}{dt^4} + R \Big( \frac{D^2x_i}{dt^2}, \frac{\partial x_i}{\partial t} \Big) \frac{\partial x_i}{\partial t}  \\&- \kappa\frac{D^2x_i}{dt^2} +\mbox{grad}_i V(x_1, x_2, ..., x_n) \Big \rangle \Bigg) dt \\&+\sum_{i=1}^n\sum_{j_i=1}^{l_i} \Big{[} \Big \langle \frac{DX_i}{dt},\frac{D^2x_i}{dt^2} \Big \rangle \\&\hspace{1.5cm}+ \Big \langle X_i, \kappa \frac{dx_i}{dt} - \frac{D^3x_i}{dt^3} \Big \rangle\Big{]}^{t_{j_i}^{-}}_{t_{j_i-1}^{+}}.\hfill\square
\end{align*}
%
%

As we are interested in variations that satisfy the boundary conditions above, the corresponding variational vector fields $X_i$ should satisfy $X_i(0)=X_i(T)=0$,  $\frac{DX_i}{dt}(0)=0$, $\frac{DX_i}{dt}(T) \in T_{x_i(T)}S$.

The next result characterizes necessary conditions for optimality in the variational collision and obstacle problem and restriction on the boundary conditions tom solve the corresponding boundary value problem.

\begin{theorem}\label{mainTh}
If $\tilde{x} \in \tilde{\Omega}$ is a local minimizer of $J$, then $ \forall i \in \{1,2...,n\}$, it satisfies
\begin{enumerate} 
\item \begin{align*}0=&\frac{D^4 x_i}{dt^4} + R \Big(\frac{D^2 x_i}{dt^2}, \frac{dx_i}{dt} \Big)\frac{dx_i}{dt}- \kappa\frac{D^2x_i}{dt^2}\\&+\mbox{grad}_i V(x_1, x_2, ..., x_n),\end{align*}
\item $x_i$ is smooth on $[0,T]$,
\item $\frac{D^2x_i}{dt^2}(T) \perp T_{x_{i}(T)}S$,
\item If the final point $x_i(T)$ is not fixed but just constrained to lie anywhere in the submanifold $S$, then  $$\kappa \frac{dx_i}{dt} - \frac{D^3x_i}{dt^3} \perp T_{x_i(T)}S.$$
\end{enumerate}
\end{theorem}
\proof
Assume $\tilde x \in \tilde\Omega$ is a local minimizer of $J$. Consider a variation of $\tilde{x}$, $\tilde{\alpha}_{r,i}(t) := (x_1(t),.. ,\alpha_{r,i}(t),... ,x_n(t))$, where $\alpha_{r,i}(t)$ is an admissible variation of $\Omega_i$ with variational vector field $X_i$. Then $\frac{d}{dr}J(\alpha_{r,i})\Big{|}_{r=0} = 0 \ \forall i \in \{1,2,...n\}$.
This is because given that the variations are independent, they have to vanish individually for each agent.

 Let us consider a variation of the $i^{th}$ agent with its variational vector field $X_i$ defined as
\begin{equation*}
f(t) \Big[ \frac{D^4x_i}{dt^4} + R\Big(\frac{D^2x}{dt^2},\frac{dx}{dt}\Big)\frac{dx}{dt} - \kappa\frac{D^2x}{dt^2}
+\mbox{grad}_i V \Big)  \Big]
\end{equation*} \\where $f:\mathbb{R}\rightarrow \mathbb{R}$ is a smooth real valued function on $[0,T]$ such that $f(t_{j_i}) = f'(t_{j_i}) = 0$ and $ f(t) > 0, t \neq t_{j_i}, j_i = 1,...,l_i$.
So, we have 
\begin{align*}
0=&\frac{d}{dr}J(\alpha_r)\Big{|}_{r=0}\\=& \int_0^T\sum_{i=1}^n  \Bigg(f(t) \Big{|}\Big{|} \frac{D^4x_i}{dt^4}+ R \Big( \frac{D^2x_i}{dt^2}, \frac{d x_i}{d t} \Big) \frac{dx_i}{dt}\\&\hspace{1.0cm}- \kappa\frac{D^2x_i}{dt^2}
+\mbox{grad}_i V(x_1, x_2, ..., x_n) \Big\|^2 \Bigg) dt \\ &\hspace{1.2cm}+ \sum_{j_i=1}^{l_i}\Big \langle f'(t)[\cdot]+f(t)\frac{D}{dt}[\cdot],\frac{D^2x_i}{dt^2} \Big \rangle \Big{]}^{t_{j_i}^{-}}_{t_{j_i-1}^{+}} \\&\hspace{1.4cm}+\sum_{j_i=1}^{l_i}\Big \langle f(t)[\cdot], \kappa \frac{dx_i}{dt} - \frac{D^3x_i}{dt^3} \Big \rangle\Big{]}^{t_{j_i}^{-}}_{t_{j_i-1}^{+}}
\end{align*}
where $[\cdot]= \frac{D^4x_i}{dt^4}+ R \Big( \frac{D^2x_i}{dt^2}, \frac{d x_i}{d t} \Big) \frac{dx_i}{dt}- \kappa\frac{D^2x_i}{dt^2}
+\mbox{grad}_i V(x_1, x_2, ..., x_n)$.

Now, since $f(t_{j_i})=f'(t_{j_i})=0$, the second and third term in the above summation vanish. 
Since $f(t)$ is greater then zero outside a set of measure zero, the first integrand has to identically vanish which establishes the first statement of the theorem, i.e.
\begin{equation*}
\begin{split}
\Big\| \frac{D^4x_i}{dt^4} +& R \Big( \frac{D^2x_i}{dt^2}, \frac{d x_i}{d t} \Big) \frac{d x_i}{d t} - \kappa\frac{D^2x_i}{dt^2}  +\mbox{grad}_i V \Big\| = 0
\end{split}
\end{equation*}
from which statement 1 follows. 

Having made the first term vanish, now, choose $X_i \in T_{x_i}\Omega_i$ such that 
\begin{align}
X_i(t_{j_i}) &= \frac{D^3x_i}{dt^3}(t_{j_i}^+) - \frac{D^3x_i}{dt^3}(t_{j_i}^-), \quad \forall j_i = 1,..,l_i-1, \nonumber\\
\frac{DX_i}{dt}(t_{j_i}) &= \frac{D^2x_i}{dt^2}(t_{j_i}^-) - \frac{D^2x_i}{dt^2}(t_{j_i}^+), \quad \forall j_i = 1,..,l_i-1,\nonumber \\
X_i(T) &= \frac{DX_i}{dt}(T) = X_i(0) = \frac{DX_i}{dt}(0)= 0.\label{constraint}
\end{align}
Therefore,
\begin{align*}
0=&\frac{d}{dr}J(\alpha_r)\Big{|}_{r=0}=\sum_{j_i=1}^{l_i-1} \Big \| \frac{D^2x_i}{dt^2}(t_{j_i}^-) - \frac{D^2x_i}{dt^2}(t_{j_i}^+) \Big \|^2  \\+& \Big \| \frac{D^3x_i}{dt^3}(t_{j_i}^+) - \frac{D^3x_i}{dt^3}(t_{j_i}^-) \Big \|^2
\end{align*}
which implies that
\begin{equation*}
\frac{D^2x_i}{dt^2}(t_{j_i}^-) = \frac{D^2x_i}{dt^2}(t_{j_i}^+) \hbox{ and } \frac{D^3x_i}{dt^3}(t_{j_i}^+) = \frac{D^3x_i}{dt^3}(t_{j_i}^-).
\end{equation*}

 Since $x_i$ is a $C^1$ curve with continuous covariant derivatives  up to order 3, $x_i$ is $C^3$ on $[0,T]$. But, we have shown that $x_i$ is the solution of a fourth order smooth ODE, which means the fourth derivative can be expressed as a smooth function of derivatives up to order 3. The $k^{th}$ order derivative can be expressed as a smooth function of derivatives upto order $(k-1)$, and so by induction, $x_i$ is smooth on $[0,T]$. Hence, statement 2 follows.
 
Now, the only non zero terms left in the cost function are given as 
\begin{align*}
   - &\Big \langle X(0),\kappa \frac {dx_i}{dt}(0)-\frac{D^3x_i}{dt^3}(0) \Big \rangle + \Big \langle \frac{DX_i}{dt}(T),\frac{D^2x_i}{dt^2}(T) \Big \rangle \\
   -&\Big \langle \frac{DX_i}{dt}(0),\frac{D^2x_i}{dt^2}(0) \Big \rangle +\Big \langle X(T),\kappa \frac {dx_i}{dt}(T)-\frac{D^3x_i}{dt^3}(T) \Big \rangle
\end{align*}

Due to the constraints \eqref{constraint}, we are always forced to choose $X_i \in T_{x_i}\Omega_i$ such that
\begin{equation*}
\begin{split}
X_i(0)=X_i(T) =\frac{DX_i}{dt}(0)= 0. 
\end{split}
\end{equation*}
Now, we choose a variational vector field $X_i$ such that we also have $\displaystyle{\frac{DX_i}{dt}(T) = \Pi_{T_{x_i(T)}S}\left(\frac{D^2x_i}{d t^2}(T)\right)}$
where $ \Pi_{T_{x_i(T)}S}V$ is the orthogonal projection onto $T_{x_i(T)}S$ of $V \in T_{x_i(T)}M$. Since
 $X_i(T) = 0$, $\displaystyle{\frac{dX_i}{dt}(T) = \frac{DX_i}{\partial t}(T) \in T_{x_i(T)}S}$. Since $X_i$ is the variational vector field of an admissible variation,
\begin{equation*}
\begin{split}
\frac{d}{dr}J(\alpha_r)\big{|}{r=0} & = \Big \langle  \Pi_{T_{x_i(T)}S}\left(\frac{D^2x_i}{\partial t^2}(T)\right), \frac{D^2x_i}{\partial t^2}(T) \Big \rangle = 0 \\
&\implies \Pi_{T_{x_i(T)}S}\left(\frac{D^2x_i}{\partial t^2}(T)\right) = 0.
\end{split}
\end{equation*} and hence statement 3 holds.

Now we consider a scenario where $x(T)$ is not fixed but is only constrained to lie at an arbitrary point in $S$. In that case $X(T)$ is not always zero. So the term in the cost function 

\begin{equation}\label{condition4}\Big \langle X_i, \kappa \frac{d x_i}{dt} - \frac{D^3\alpha}{\partial t^3} \Big \rangle(T)\end{equation} in the expression for $\frac{dJ}{dr}$ and it does not vanish because $X_i(T)=0$.
Since $x_i(T)\in S$, we must now have $X_i(T)\in T_{x_i(T)}S$. By choosing a variation such that $$X_i(T)=\Pi_{T_{x_i(T)}S}\left(\kappa \frac{dx_i}{dt} - \frac{D^3 x_i}{dt^3} \right)(T),$$ it follows that $$\kappa \frac{dx_i}{dt} - \frac{D^3x_i}{dt^3} \perp T_{x_i(T)}S$$ because if otherwise, it will contribute a strictly positive value for $\frac{d}{dr}J(\alpha_r)|_{r=0}$ when $X_i$ is chosen such that Hence statement $4$ holds. \hfill$\square$

\vspace{.5cm}

\textbf{Boundary conditions and well-posedness:}
Note that condition (1) in Theorem 3.2 is a fourth order ODE in each of the $n$ $x_i$ variables which are $s$ dimensional. So, the total order of the system of ODEs is $4sn$, and hence, for well posedness, $4sn$ conditions are required in the form of initial and/or boundary conditions which is verified for each case in the tables given below. Recall again that $S$ is a $m$ dimensional manifold. 
\begin{itemize}
    \item Case 1: Final position fixed, final velocity in $TS$ \\ \\ 
        \begin{tabular}{l|c}
        \hline
      \textbf{Condition} & \textbf{$\#$ of constraints}\\
        \hline
        $x_i(0)=p_0^i$ & $sn$\\
        
        $\frac{dx_i}{dt}(0)=v_0^i$ & $sn$ \\
        
        $x_i(T)=p_T^i\in S$ & $sn$ \\
       
        $\frac{dx_i}{dt}(T)\in T_{x_i(T)}S$ & $n(s-m)$ \\ 
        
        Condition (3) in Theorem 3.2 & $mn$ \\
        \hline
        TOTAL &    $4sn$

	\end{tabular} \\
	\item Case 2: Final velocity also fixed \\ \\ 
	 \begin{tabular}{l|c}
        \hline
      \textbf{Condition} & \textbf{$\#$ of constraints}\\
        \hline
        $x_i(0)=p_0^i$ & $sn$\\
        
        $\frac{dx_i}{dt}(0)=v_0^i$ & $sn$ \\
       
        $x_i(T)=p_T^i$ & $sn$ \\
       
        $\frac{dx_i}{dt}(T)=v_T^i$ & $sn$ \\ 
        
        Condition (3) in Theorem 3.2 & Vacuous \\
      \hline
        TOTAL &    $4sn$\\

	\end{tabular} \\
	\item Case 3: Final position and velocity not fixed but in $TS$\\ \\
	 \begin{tabular}{l|c}
        \hline
      \textbf{Condition} & \textbf{$\#$ of constraints}\\
        \hline
        $x_i(0)=p_0^i$ & $sn$\\
       
        $\frac{dx_i}{dt}(0)=v_0^i$ & $sn$ \\
        
        $x_i(T)\in S$ & $s(n-m)$ \\
       
        $\frac{dx_i}{dt}(T)\in T_{x_i(T)}S$ & $s(n-m)$ \\ 
      
       Condition (3) in Theorem 3.2 & $mn$ \\
      
        Condition (4) in Theorem 3.2 & $mn$ \\
        \hline
        TOTAL &    $4sn$\\

	\end{tabular}
\end{itemize}

\section{Examples}
\subsection{Application to variational obstacle avoidance problem for a planar rigid body on $SE(2)$.}
The  special euclidean Lie group $SE(2)$ consists of all the transformations of $\mathbb{R}^2$ of the form $z \mapsto Rz+v$, where $v\in \mathbb{R}^2$ and $R\in SO(2)$. This Lie group is, as a smooth manifold, diffeomorphic to $\mathbb{R}^{2}\times S^{1}$.




The Riemannian metric on $SE(2)\simeq\mathbb{R}^{2}\times \mathbb{S}^{1}$, locally parametrized by $\gamma=(q,\theta)=(x,y,\theta)$, is determined by a diagonal matrix with $\mbox{diag =}(m,m,J)$. The reason stems from the fact that it is the same matrix that yields the kinetic energy of a planar rigid body with $m$ being the mass of the body and $J$ the moment of inertia of the body. The curvature tensor is zero under this metric.

We denote by $\gamma_i=(q_i,\theta_i)=(x_i, y_i,\theta_i)$ the trajectory of the $i$-th agent, $i=1,2,..., n$, and represent by $p^k=(p_x^k,p_y^k)$ the center of an obstacle with circular shape in the $xy$-plane, with radius $r^k$. The submanifold $S$ of $SE(2)$ given by $\mathbb{S}^1\times S(p^0,l)$, where $S(p^0,l)$ is the circle with center $p^0\in\R^2$ and radius $l$.
We consider the artificial potential function $V$ given by  
\begin{align*}V(\gamma) =& \frac{1}{2}\sum_{i=1}^n\sum_{j=1}^{n}\frac{\sigma}{\|q_i - q_j\|^2-4d^2}\\&+\sum_{k=1}^{\alpha}\sum_{i=1}^n\frac{\tau}{\|q_i - p^k\|^2-(d+r^k)^2},\end{align*} where $d$ is the radius of the smallest ball containing each agent and $\tau,\sigma\in\mathbb{R}^{+}$. 
Note that, for $i=1,2,..,n$, 
\begin{align*}\frac{\partial V}{\partial x_i} (\gamma)=&\sum_{j=1}^{n}\sum_{j=1}^{n}\frac{\sigma}{(x_i-x_j)^2+(y_i-y_j)^2-4d^2}\\&\times((x_j-x_i)\frac{\partial}{\partial x}+(y_j-y_i)\frac{\partial}{\partial y})\\
&+\sum_{i=1}^{n}\sum_{k=1}^{\alpha}\frac{2\tau}{(x_i-p_x^k)^2+(y_i-p_y^k)^2-(d+r^k)^2}\\&\times((p_x^k-x_i)\frac{\partial}{\partial x}+(p_y^k-y_i)\frac{\partial}{\partial y}).
\end{align*}

By Theorem \ref{mainTh}, the equations determining necessary conditions for the extrema in the variational collision and obstacle avoidance problem are\begin{align*}
                   \theta_i^{(4)}=&\kappa\theta_i'',\\
                  x_i^{(4)}=&\kappa x_i''+\sum_{i=1}^{n}\sum_{j=1}^n\frac{\sigma(x_j-x_i)}{(x_i-x_j)^2+(y_i-y_j)^2-4d^2}\\+&\sum_{i=1}^{n}\sum_{k=1}^{\alpha}\frac{2\tau (p_x^k-x_i)}{(x_i-p_x^k)^2+(y_i-p_y^k)^2-(d+r^k)^2},\\
                  y_i^{(4)}=&\kappa y_i''+\sum_{i=1}^n,\sum_{j=1}^n\frac{\sigma(y_j-y_i)}{(x_i-x_j)^2+(y_i-y_j)^2-4d^2}\\&+\sum_{i=1}^{n}\sum_{k=1}^{\alpha}\frac{2\tau (p_y^k-y_i)}{(x_i-p_x^k)^2+(y_i-p_y^k)^2-(d+r^k)^2},
                 \end{align*}  with given beginning position and velocity
             $(x(0),y(0),\theta(0))$ and   $(x^{\prime}(0),y^{\prime}(0),\theta^{\prime}(0))$ and satisfying the end-boundary conditions $x_i''(T)(x_i(T)-p_x^0)+y_i''(T)(y_i(T)-p_y^0)=0$ and $(\kappa x_i'(T) - x_i''''(T))((x_i(T)-p_x^0)+(\kappa y_i'(T) - y_i''''(T))((y_i(T)-p_y^0)=0$.

\subsection{Application to variational collision and obstacle avoidance problem on $S^2$.}
Next, we study the variational collision and obstacle avoidance problem on the sphere. This problem has applications in, for instance, trajectory planning for aircrafts in the presence of unsafe regions of flight (which can be considered as obstacles). Here, $S^2=\{x \in \mathbb{R}^3, ||x||_2=1\}$ is considered as a Riemannian submanifold of $\mathbb{R}^3$ and hence, its Riemannian metric is inherited from the standard metric on $\mathbb{R}^3$ as an inner product space. $S^2$ can be parametrised in spherical polar coordinates (or longitudes and latitudes) as  $(x_1,x_2,x_3)(\theta,\phi)=(\sin\theta \sin\phi, \sin\theta \cos\phi, \cos\theta)$ and this serves as a coordinate system for $S^2$ except at a great circle passing through $(0,0,1)$ where the parametrization becomes singular and/or non-injective. 
With this choice of parametrization, if $x(t)=(\theta(t),\phi(t))$ is the representation of a parametrized curve (in this chart), then it can be shown that for the actual curve $x(t)$,
\begin{align*}
    \frac{D^4x}{Dt^4}&=(\theta''''  +5\sin(2\theta)\theta'^2\phi'^2+(1-7\cos^2\theta)\theta''\phi'^2\\&+(5-17\cos^2\theta \theta'\phi'\phi''-3\sin\theta \cos\theta \phi''^2\\&-2\sin(2\theta)\phi'\phi'''+\sin\theta \cos^3\theta \phi'^4)\frac{\partial}{\partial \theta}+\\&(\phi''''-7\theta'\theta''\phi'-5\theta'^2\phi'')+4cot\theta \theta''' \phi' + (6cot\theta \theta'' \phi'' \\&+4\cot\theta \phi'''\theta'+(\sin(2\theta)-\cot\theta(5\cos^2\theta-1))\theta'\phi'^3)\\&-6\cos^2\theta\phi'^2\phi''-2\cot\theta\phi'\theta'^3\frac{\partial}{\partial \phi}
\end{align*}
 and
 $R(\frac{D^2x}{Dt^2},\frac{dx}{dt})\frac{dx}{dt}=0$.
 
Let us assume $\kappa=0$ and there are only two agents indexed by $i\in\{1,2\}$. 
 Next, we need a potential function for each agent to capture the obstacle avoidance terms. 
 
 Let the obstacle be described by  $\{x\in S^2 | f(x)\leq0\}$ where $f$ is a smooth real valued function on $S^2$. Assume thatn the obstacle is a circle centered at the north pole in $S^2$ of geodesic radius $\frac{\pi}{4}$, then we have in local coordinates, $f(\theta,\phi)=\theta-\frac{\pi}{4}$. Then we choose a smooth potential function that is non-negative for $f>0$ which approaches to infinity as $f\rightarrow 0^+$. It should also become constant when $f$ is sufficiently large and positive. For the above situation, a  candidate $V$ would be $$V_i(\theta_i,\phi_i)=\frac{1}{(\theta_i-\frac{\pi}{4})^2}.$$ 
 
 Then, using the Riemannian metric of $S^2$ in the above chart, we can find the gradient of $V_i$ which is required in Theorem 3.2. It is given by $$\hbox{grad}(V_i(\theta_i,\phi_i))=\Big{[}0,...,\frac{-2}{(\theta-\frac{\pi}{4})^3},0,...\Big{]}^T$$ where the non-zero pair is the $i^{th}$ pair.
  
 Now, we need another potential function to capture the collision avoidance terms. As far as the collision term is concerned, if $d_M(x,y)$ denotes the Riemannian distance between two points $x,y$, for the sphere it may be defined by using the exponential function as $$d_{S^1}(x,y)=||\exp_x^{-1}y||,$$ and it can be shown that $$\frac{\partial}{\partial r}d_{S^1}(x,y(r))\big|_{r=0}=-\Big{<}\frac{\partial y}{\partial r}\big|_{r=0}, \exp_{y(0)}^{-1}x\Big{>}.$$
  Then we need a potential function that goes to infinity as the inter-agent distance approaches to zero and it is close to zero when the inter-agent distances are large enough. A candidate function would be 
 $\displaystyle{F(d^2_S)=\frac{1}{2}\frac{1}{d_S^2}}$. 
 Then, it can be shown as in [20] that for the potential function $V_{12}=F(d_S^2(x_1,x_2))$, $$(\hbox{grad} V)^i=(-F'(d_S^2)\exp_{x_1}^{-1}x_2)$$ where$(\hbox{grad} V)^i$ denotes the $i^{th}$ component of gradient. 
 
 Now for two points $x,y\in S^2$, we have 
\begin{align*}
    \exp_x^{-1}(y)=\cos^{-1}\bigg(\big<x,y\big>\frac{y-\big<x,y\big>x}{\sqrt{1-\big<x,y\big>^2}} \bigg) 
\end{align*} and hence $\displaystyle{d_S(x,y)=\bigg|\bigg| \exp_x^{-1}(y)\bigg|\bigg|=\cos^{-1}(\big<x,y\big>)}$. 
So the final potential function is constructed by combining the obstacle and collision terms as $V=V_1+V_2+V_{12}$, and therefore substituting the respective gradients. \\ \\
 So, equipped with a parametric representation of the curve, the conditions of Theorem 3.2 give a two point boundary value problem for the parametrized curve $(\theta_i(t),\phi_i(t))$ may be solved numerically to determine the optimal trajectories.
\section{Conclusions and Future work}

We discussed the problem of collision and obstacle avoidance of multi-agent systems on a  Riemannian manifold and
derived, from the point of view of calculus of variations,
first order necessary conditions for optimality in the
problem. We have shown how the main result can be applied
for the particular cases of planar rigid bodies and agents evolving on a sphere. In future work we intend to extend the results presented in this paper to the Lie group case and sub-Riemannian problems which allows to consider non-holonomic vehicles.
as well as to explore numerical results.



\end{document}